# Kohn Anomaly in $MgB_2$ by Inelastic X-Ray Scattering

A.Q.R. Baron[1], H. Uchiyama[2], Y. Tanaka[3], S. Tsutsui[1], D. Ishikawa[3*],
S. Lee[2], R. Heid[4], K.-P. Bohnen[4], S. Tajima[2] and T. Ishikawa[1,3].

[1]SPring-8/JASRI, 1-1-1 Kouto, Mikazuki-cho, Sayo-gun, Hyogo, 679-5198, JAPAN

[2]Superconductivity Research Laboratory, ISTEC, Tokyo, 135-0062, JAPAN

[3]SPring-8/RIKEN, 1-1-1 Kouto, Mikazuki-cho, Sayo-gun, Hyogo, 679-5148, JAPAN

[4]Forschungszentrum Karlsruhe, Institut für Festkörperphysik, P.O.B. 3640, D-76021, GERMANY

We study phonons in $MgB_2$ using inelastic x-ray scattering (1.6 and 6 meV resolution). Our measurements show excellent agreement with theory for the dispersion and line-width: we clearly observe the softening and broadening of the crucial $E_{2g}$ mode through the Kohn anomaly along ΓM. Low temperature measurements (just above and below $T_c$) show negligible changes for the momentum-transfers investigated, and no change in the $E_{2g}$ mode at A between room temperature and 16K. We report the presence of a longitudinal mode along ΓA near in energy to the $E_{2g}$ mode that is not predicted by theory.

MgB$_2$ has generated an immense amount of interest since its high T$_c$ ~39K was first demonstrated two years ago [1]. This is the highest critical temperature yet achieved for a simple metallic material, and rather outside the range for T$_c$ using standard estimates. Its simple structure, with only three atoms/unit cell, and without magnetism, makes calculations relatively tractable, allowing the careful comparison of theory with experiment. Based on a wealth of experimental and theoretical work, MgB$_2$ is now understood to be a phonon-mediated Eliashberg superconductor with multiple gaps [2-7]. The presence of a relatively big (~7 meV) isotropic gap on the sigma hole Fermi surfaces dominates the superconducting properties and is largely responsible for the unusually high T$_c$.

Investigation of the phonons in MgB$_2$ is important as they play a crucial role in its superconductivity. Several calculations of phonon dispersion and electron-phonon coupling have appeared [8-10] and [6,7], and give mostly consistent results. The large gap on the sigma Fermi surface is the result of extremely strong electron-phonon coupling (EPC) of this surface to the basal-plane boron E$_{2g}$ mode. While remaining within the Eliashberg model, the strong coupling to a single mode is very unusual, and means the investigation of this phonon mode is of particular interest. Experimentally, the EPC should result in both a large phonon line-width and, in the neighborhood of 2k$_F$, in a steep change in phonon mode frequency with phonon momentum – essentially a Kohn anomaly [11]. Even more extreme cases of a Kohn anomaly have been considered in related materials [12]. Earlier experimental work to probe phonon structure in MgB$_2$ includes neutron density of states measurements on polycrystalline samples [13] and Raman scattering data [14]. The first is an integral method which does not easily allow investigation of specific modes, while the second is limited to probing phonons (and in this case only the E$_{2g}$ mode) at Γ.

The method of choice for probing phonons in MgB$_2$ at non-zero wave vectors is inelastic x-ray scattering (IXS). This is primarily because presently available single crystals are too small (~0.01 mm$^3$) for inelastic neutron scattering measurements – the more traditional way of measuring phonon dispersion in materials. Over the last few decades, improvements in x-ray sources and optics have been dramatic, and, especially, the advent of "third" generation synchrotron radiation facilities means that it is now not too difficult to generate an Å wavelength x-ray beam of size ~100 μm, bandwidth ~meV and intensity ~5 GHz/meV. Given an IXS spectrometer operating at a third generation source, experiments with ~100 μm size samples are then relatively straightforward. Thus, recently, IXS measurements on MgB$_2$ were presented [15] using an established spectrometer operating at the European Synchrotron Radiation Facility (ESRF) in France. The work reported here was begun before the report of the ESRF work, during the commissioning of the newest operating high resolution IXS spectrometer at a third generation source, SPring-8 in Japan. In comparison to [15], we provide significant new information about the phonons in MgB$_2$. In general, and like [15], our dispersion results show good agreement with calculation. However, in addition, we directly measure changes in the line width of the crucial E$_{2g}$ mode through the Kohn anomaly resulting from the coupling of this mode to the sigma sheets of the Fermi surface. We provide temperature dependent data (whereas that in [15] is at room temperature). Finally we show the presence of an anomalous optical mode, similar in energy to that of the $E_{2g}$ mode, but having a different line-width and different symmetry.

This work was carried out using the high resolution inelastic x-ray scattering spectrometer at BL35XU of SPring-8 [16]. A backscattering geometry ($\pi/2 - \Theta_B \leq 0.3$ mrad) is used for both monochromator and analyzer crystals in order to obtain large angular acceptance while preserving excellent energy resolution. High resolution data was collected using the Si (11 11 11) reflection at 21.747 keV, providing $\sim 3 \times 10^9$ photons/s (100 mA electron beam current) in a 0.8 meV bandwidth onto the sample and an over-all resolution of 1.6 to 1.8 meV (depending on the analyzer crystal). A grazing incidence geometry for the backscattering monochromator was crucial to avoid heat-load broadening of the monochromator response function. We also used a high flux setup, silicon (888) reflection, 15.816 keV, to investigate the weaker optical modes, especially the $E_{2g}$. This setup provided $\sim 3 \times 10^{10}$/photons/sec in a 4 meV bandwidth onto the sample, with an over-all resolution of 6.0 to 6.2 meV. The use of 4 analyzers crystals, placed with 0.78 degree spacing on the 10m two-theta arm (horizontal scattering plane), and 4 independent detectors (room temperature CdZnTe chips, with dark count rates ~0.001Hz at 22 keV and 0.002 Hz at 16 keV), allowed collection of 4 momentum transfers simultaneously. This greatly facilitates data collection for longitudinal modes where all 4 analyzers may be placed along a common symmetry direction. The full 95 mm diameter of each analyzer crystal was used to get a maximum count rate, so the momentum resolution was ~0.09 (0.07) $\text{Å}^{-1}$, full width, at the Si (11 11 11) (Si(8 8 8)) (most measurements made at scattering angles between 30 and 50 degrees). The beam size at the sample depended on the setup, varying from 70x80 $\mu m^2$ (vertical x horizontal) in the FWHM (full width at half maximum) at the (11 11 11) to 130 x 100 $\mu m^2$ at the (888). For all work, a slit before the sample insured proper alignment (beam onto the sample, and into the center of the spectrometer) and spectra were normalized using the intensity measured downstream of this slit.

Two different $MgB_2$ samples were investigated, each a thin platelet (~50-70 μm thick) in the c-axis direction, with transverse dimensions ~ 0.2 x 0.5 $mm^2$ (though the shape was not regular), grown according to [17]. $T_c$ for the samples was >38 K with a transition region of ≤0.3K. Typical rocking curve widths were 0.03 to 0.07 deg. for lower order reflections. The crystals were mounted on thin rods attached to the copper cold-finger of a closed cycle cryostat with the sample c axis normal to the common rod-refrigerator axis. This orientation provided maximum flexibility for the spectrometer to investigate various momentum transfers and phonon polarizations, given the constraints imposed by the geometry of the Eulerian cradle and the attachments for the refrigerator. With these samples, data collection times varied from about 2 hours for quick scans of acoustic modes, to about 8 hours as the typical scan time for 6 meV resolution data including the $E_{2g}$ mode (e.g. fig. 1), to about 20 hours for a couple scans with high resolution and weaker modes.

Figure 1 shows the data measured along ΓM, while Fig. 2 shows the dispersion determined by fitting the measured spectra. Even at the level of the raw data in Fig. 1, the softening and the broadening of the $E_{2g}$ mode as one approaches Γ are readily apparent. (Note that while the two $E_{2g}$ modes are not degenerate along ΓM – see Fig. 2 - our scattering geometry was chosen so that only one mode was excited). Considering this as the Kohn anomaly expected from the coupling of the $E_{2g}$ mode to the sigma Fermi surfaces, one then sees nearly what one naïvely expects. In detail, there are two sigma surfaces, each approximately cylindrical with axis parallel to ΓA, one having a slightly larger diameter than the other [4] with average radii of 0.17 and 0.25 Å$^{-1}$ (0.16 and 0.21 ΓM) [18]. Thus, as one goes from smaller phonon momenta ($q<2k_F$) which can promote electrons from filled states to un-filled states across these surfaces, to larger momenta which bridge the entire surface and so can not excite electrons (as this would shift them from filled states to filled states), one sees a narrowing of the phonon mode corresponding to an increased lifetime. Changes in the phonon linewidth (essentially the imaginary part of the phonon self energy) are accompanied by changes in the mode energy (real part of the self energy) [19], which generally shows mode softening in the regions where there is more coupling – this can be considered the result of increased screening in the region of larger EPC[11][12]. When one then considers that the Fermi-surface has some dispersion, so that $k_F$ varies along ΓA [4][18], the relatively smooth variation in phonon parameters (as opposed to the sharper Kohn anomalies in other materials) becomes very reasonable.

Quantitative results were obtained by fitting the spectra with a sum of damped harmonic oscillators (DHOs) [20] for the phonon modes. Mode intensities were scaled by a detailed balance factor and then a Lorentzian (line width much less than the resolution) was added to account for the elastic diffuse scattering (resulting from, e.g., crystal imperfections and surface scattering). The resolution function (as determined for each analyzer using scattering from PMMA (poly methyl methacrylate)) was then convolved with the calculation and the parameters (mode energies, widths and amplitudes) were optimized by a least-squares fit to the data. Initially, only modes expected from the calculations were included. However, this led, in some cases, to poor fits in the lower energy region between the Stokes and anti-Stokes lines of the acoustic modes. Thus a broad low-energy component was included (modeled by a DHO of line width > 10 meV, and center < 35 meV ) to improve the fit quality. This made negligible changes in the fitted mode energies, and a consistent, but small, reduction in line width as the fits to the tails of the modes improved. In general, errors in mode energies from fitting were small compared with point size, excepting the $E_{2g}$ mode along ΓA (not plotted – see the discussion below) and the very weak mode near the $B_{1g}$ energy along ΓA (as plotted). The variation from point to point over several different measurements is probably the best indication of absolute uncertainty in mode energy. In general, the agreement between our measured dispersion and the calculations of [9] can be seen to be excellent, excepting the anomalous intensity along ΓA (discussed below).

Returning to the Kohn anomaly along ΓM, we compare the $E_{2g}$ line width more precisely with theory. The width (lifetime) is a direct measure of the decay modes, and in this case, where anharmonic contributions are expected to be small [15], should be directly related to the electron-phonon coupling as given in [19]. Calculations were performed similarly to [9] using a mixed-basis pseudopotential method, combining plane waves and local functions for the valence states (see [12] for details). Harmonic phonons and screened electron-phonon matrix elements were calculated within the mixed-basis perturbation approach [22]. In order to get accurate results, dynamical matrices were obtained on a hexagonal (18x18x6) q-point mesh. The screened EPC matrix element is directly accessible from quantities obtained in the calculation of the dynamical matrix. The Fermi-surface average was approximated by a sum over a dense ($36^3$) k-point mesh. The line width (after correction for instrumental resolution) measured along ΓM is compared with this calculation in Fig. 3, with good agreement, providing crucial confirmation of the picture of the phonon role in the superconductivity of $MgB_2$.

Measurements of the $E_{2g}$ mode were also made along ΓA in a transverse geometry (similar to [15]) as seen Fig. 4. However, while the $E_{2g}$ mode was clearly observed, determining the mode energy and width is difficult: fits could be made with line widths varying between 25 and 40 meV, depending on the parameters chosen for the other lines, the line shape chosen for the $E_{2g}$ mode (Lorentzian or DHO), and the $E_{2g}$ mode energy. Thus, noting that there is some uncertainty in the correct line shape to use (the available calculations of the effects of EPC generally assume that the broadening is much less than the line width), the only statement that we make with confidence is that the line width is certainly large, probably > 25 meV.

The temperature dependence of the **$E_{2g}$** mode is of some interest. Raman measurements (at Γ) suggested some change in width when crossing $T_c$ [14], while one calculation [6] suggested there should be a shift in frequency. The last was surprising in that while modification of phonon structure near $T_c$ is well known [e.g. 22] for modes with energies in the neighborhood of twice the gap energy, 2Δ (~15 meV in this case), they are unusual for high energy modes. Figure 4 shows measurements at the A point on one sample as a function of temperature. In short, at the level of the data here, there is no evidence for any change on crossing $T_c$ – and indeed no change at all in the $E_{2g}$ mode between room temperature and 16K. This is consistent with recent calculations of the anharmonic frequency shift [24]. We note that the normalization of the data in figure 4 is only to monitor counts – there are no relative scale factors between the spectra – so this speaks well for the stability and reproducibility of the spectrometer, even with a small sample.

Finally, we consider the presence of anomalous intensity along ΓA (Fig. 2). It is anomalous because it appears in the purely longitudinal (0 0 3+ξ) geometry where both the $E_{2g}$ and $B_{1g}$ are forbidden: for the $E_{2g}$, the motions are purely within the boron plane, so have no component along ΓA, while the $B_{1g}$ involves equal and opposite motions of adjacent boron atoms out of the plane, so has no net contribution along ΓA. However, this intensity (especially the stronger mode at 65 meV) appears in two setups, with different samples, different x-ray energy and different resolution, so seems a solid result. In addition, the line width of the 65 meV mode is consistently 3 to 5 meV along ΓA, in disagreement with the very much larger $E_{2g}$ line width expected and measured (Fig. 4) in a transverse geometry. One possibility, as this is an x-ray scattering measurement, is that this intensity could be due to some electronic excitation. However extensive simulation of the scattering (actually in regard to the possibility of directly seeing scattering from the superconducting gap [25]) using approximate (tight binding) models of the band structure based on [4][5][18], did not provide any likely candidates. This intensity could be due to details of the sample structure that are not included in the model calculations. In particular, there has been some suggestion that there is a Mg deficiency (4-5%) in samples prepared in the same fashion as those used here [26]: the flat dispersion of the stronger mode observed here would, for example, be compatible with Einstein type behavior of a localized defect mode.

In sum, this work provides a good picture of the phonon dispersion and line width variation in $MgB_2$. Naïve expectations for the $E_{2g}$ mode behavior through the Kohn anomaly region are well born out by experiment, while agreement with detailed calculations provides crucial confirmation of the understanding of the superconductivity in $MgB_2$. The small temperature dependence of the $E_{2g}$ mode and other modes near $T_c$, while not surprising based on previous work with superconductors, is important in the context of discussion of anharmonicity in $MgB_2$, setting some limits for calculations. Anomalous intensity along ΓA remains un-explained in detail, but might be a localized defect mode. Similar considerations (sample disorder) might also explain the low energy background needed to get good fits to some of the data.

A.B. is grateful to P. Johansson for discussion and help with the electronic scattering calculations. We thank Y. Endoh for his interest in this work. This work was carried out at SPring-8 under proposals 2002A0559 and 2002B0594. H.U. acknowledges support by the JSPS Research Fellowship for Young Scientists. This work was partially supported (through ISTEC-SRL) by the New Energy and Industrial Technology Department Organization (NEDO) as Collaborative Research and Development of Fundamental Technologies for Superconductivity Applications.

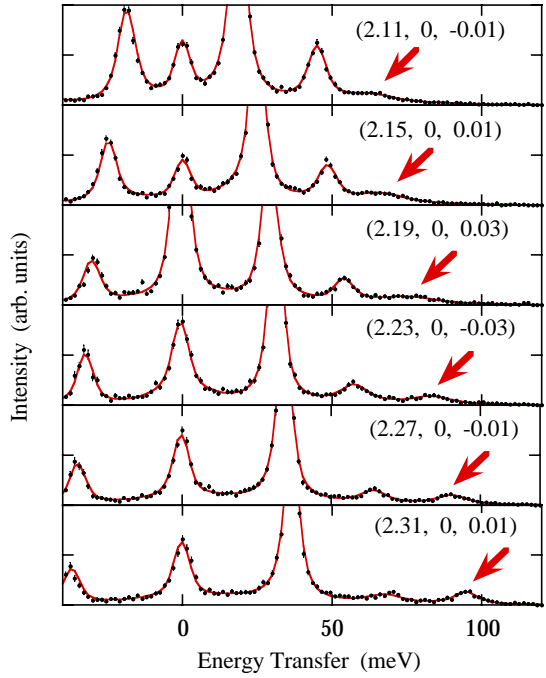

Figure 1. **Spectra along ΓM.** (Room temperature and 6 meV resolution). Note the shift and increasing line width of the $E_{2g}$ mode (indicated by arrows). Solid lines are fits to the data.

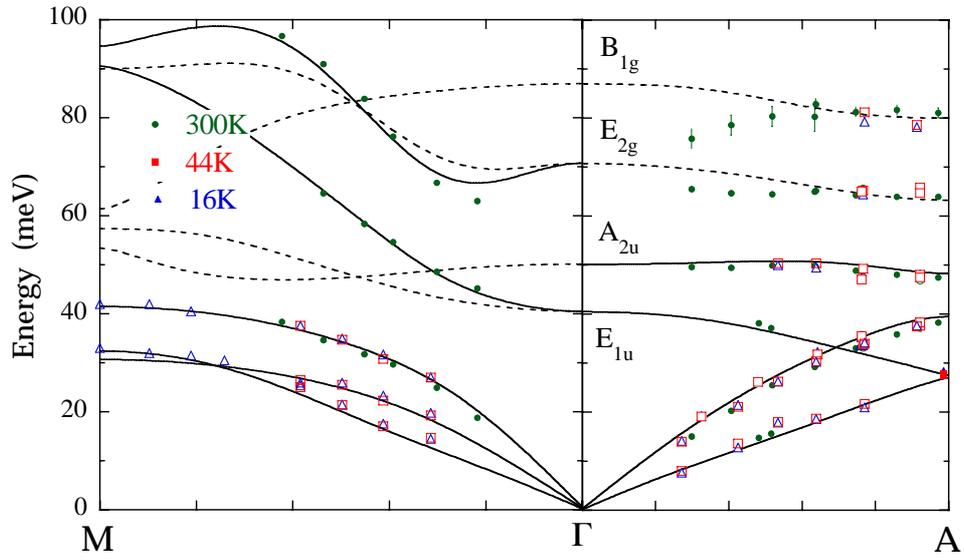

Figure 2. **Phonon dispersion**. Points show measured values while lines are calculations of [9]. Dashed lines show branches not investigated in this work due to polarization selection rules or choice of scan range. Open symbols are measured using high resolution, while filled symbols are measured using a high flux setup. Circles (green, online) are room temperature measurements, squares (red, online) at 44 or 43 K, and triangles (blue, online) at 16K. Points approximately at the calculated $E_{2g}$ and $B_{1g}$ energies along $\Gamma A$ are anomalous longitudinal modes. See text for discussion.

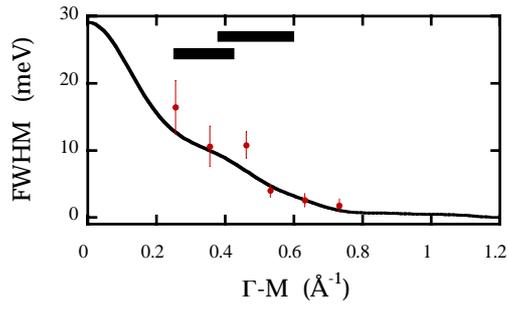

Figure 3. **Measured and calculated E$_{2g}$ line width along ΓM (M at 1.17 Å$^{-1}$).** See text for details. The horizontal bars on the figure show the diameters of the sigma Fermi surfaces projected into the plane perpendicular to the ΓA axis (from [18]) giving a naïve indication of the region where the Kohn anomaly should occur.

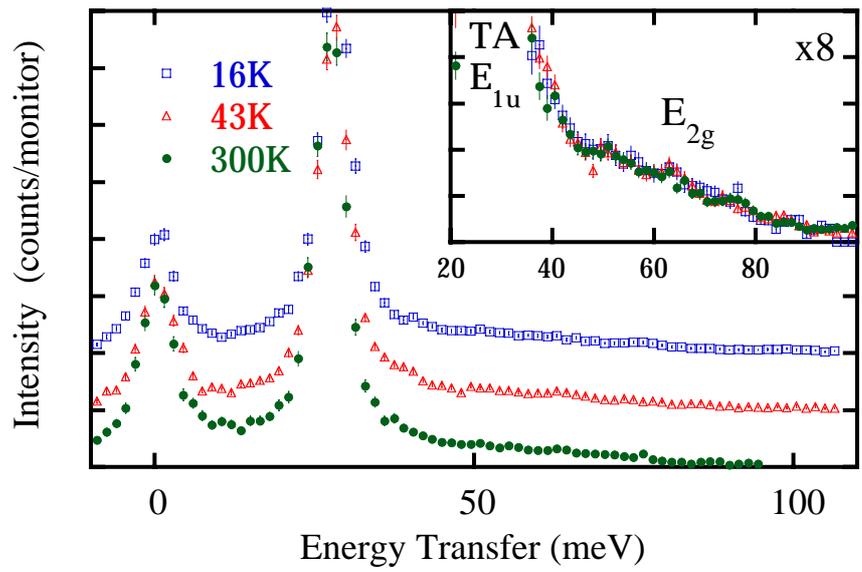

Figure 4. **Temperature dependence of the transverse modes at A (2 1 0.5).** 6 meV resolution, offset for clarity in the main figure, but overlapped in the inset. See text for discussion.